\documentstyle[12pt,epsf]{article}
\newcommand{\beq}{\begin{equation}}
\newcommand{\eeq}{\end{equation}}
\newcommand{\beqa}{\begin{eqnarray}}
\newcommand{\eeqa}{\end{eqnarray}}
\def\simgt{\rlap{\lower 4 pt \hbox{$\mathchar \sim$}} \raise 1pt \hbox {$>$}}
\def\simlt{\rlap{\lower 4 pt \hbox{$\mathchar \sim$}} \raise 1pt \hbox {$<$}}

\newcommand{\be}{\begin{equation}}
\newcommand{\ee}{\end{equation}}
\newcommand{\bea}{\begin{eqnarray}}
\newcommand{\eea}{\end{eqnarray}}

\newcommand{\gm}{\gamma}
\newcommand{\Gm}{\Gamma}

\newcommand{\eps}{\epsilon}

\catcode`\@=11
\def\@citex[#1]#2{\if@filesw\immediate\write\@auxout{\string\citation{#2}}\fi
  \def\@citea{}\@cite{\@for\@citeb:=#2\do
    {\@citea\def\@citea{,\penalty\@m}\@ifundefined
       {b@\@citeb}{{\bf ?}\@warning
       {Citation `\@citeb' on page \thepage \space undefined}}%
\hbox{\csname b@\@citeb\endcsname}}}{#1}}
\def\citer{\@ifnextchar [{\@tempswatrue\@citexr}{\@tempswafalse\@citexr[]}}
\def\@citexr[#1]#2{\if@filesw\immediate\write\@auxout{\string\citation{#2}}\fi
  \def\@citea{}\@cite{\@for\@citeb:=#2\do
    {\@citea\def\@citea{--\penalty\@m}\@ifundefined
       {b@\@citeb}{{\bf ?}\@warning
       {Citation `\@citeb' on page \thepage \space undefined}}%
\hbox{\csname b@\@citeb\endcsname}}}{#1}}
\begin{document}

\begin{titlepage}
\begin{flushright}
        CERN-TH/97-315\\
        hep-ph/9711391\\ 
        November 1997\\
\end{flushright}
\vskip 1.8cm
\begin{center}
 \boldmath
{\Large\bf Asymptotic expansion of Feynman integrals\\[0.2cm]
 near threshold} \unboldmath
\vskip 1.2cm
{\sc M. Beneke}
\vskip .3cm
{\it Theory Division, CERN, CH-1211 Geneva 23}
\vskip .7cm
{\sc V.A. Smirnov}
\vskip .3cm
{\em Nuclear Physics Institute, Moscow State University, \\
119889 Moscow, Russia}
\vskip 1.0cm
\end{center}

\begin{abstract}
\noindent We present general prescriptions for the asymptotic expansion 
of massive multi-loop Feynman integrals near threshold. 
As in the case of previously known prescriptions for 
various limits of momenta and masses, 
the terms of the threshold expansion are associated with subgraphs
of a given graph and are explicitly written through Taylor expansions of the
corresponding integrands in certain sets of parameters. They are
manifestly homogeneous in the threshold expansion parameter, 
so that the calculation of 
the given Feynman integral near the threshold reduces to the calculation 
of integrals of a much simpler type. The general 
method is illustrated by two-loop two-point and three-point diagrams. 
We discuss the use of the threshold expansion for problems 
of physical interest, such as the next-to-next-to-leading order  
heavy quark production cross sections close to threshold and matching 
calculations and power counting 
in non-relativistic effective theories. 
\end{abstract}

\vfill

\end{titlepage}


{\large\bf \noindent 1. Introduction}\\
 
\noindent Many interesting processes in particle physics, 
in particular those in which heavy quarks participate, involve more than 
one mass scale. Such processes are notoriously difficult to calculate 
in perturbation theory beyond the one-loop level. To proceed one has to 
resort to approximations, either numerical or analytical. 
Among the latter, asymptotic expansions in certain ratios of mass 
scales appear most promising, because the analytic complexity 
is substituted by the algebraic complexity associated with obtaining 
a large number of terms in the expansion, which, however, can be 
delegated to a computer. To achieve this goal, the integrals that 
appear in the calculation of any given term in the expansion should 
(of course) be simpler than the original Feynman diagram. 
In particular, this means that 
the expansion should be manifestly homogeneous, that is, every integral 
that appears in the construction should contribute only to a single 
power in the expansion parameter. 

Let us consider, for the purpose of discussion, a quantity that 
depends on a single kinematic invariant $q^2$ and a particle mass $m$, 
such as the two-point functions of heavy quark currents, or 
the production cross section for a pair of heavy quarks. There exist 
general explicit prescriptions \cite{Go,Ch,Sm1} (see \cite{Sm2} for 
brief reviews) to obtain the asymptotic expansion of these (and other) 
quantities as $q^2\to\infty$ or $m\to\infty$.  
In this paper we propose a 
prescription to obtain the asymptotic expansion close to threshold, 
that is, as $q^2\to 4 m^2$. This limit has not been explored 
systematically yet, although it is of considerable interest for 
a field theoretical description of non-relativistic systems.  
We illustrate our method through examples in 
Sects.~2 and 4. In these examples we consider only scalar 
propagators. The extension to fermion and gauge field propagators 
complicates the numerators of the integrals, but is 
straightforward methodically. The general 
structure of the expansion, which we 
formalize in Sect.~3, follows the same strategy used to obtain the 
the expansions in limits of large/small momenta and masses, 
although the threshold expansion is quite different, in particular 
because we have to deal with three 
different scales near threshold. The method obviously generalizes to 
the threshold production of two unequal-mass particles and in fact 
to any particle threshold, when some massive particles are slow, 
although we do not treat these cases in this paper. Physically, the 
mathematical problem of constructing an asymptotic expansion is 
closely related to the notion of effective field theories. For example, 
the homogeneity property mentioned above translates into the 
property of manifest power counting for the effective Lagrangian. 
The threshold expansion provides some insight into how to construct a 
non-relativistic effective field theory within dimensional regularization, 
an issue that has received some attention recently. We discuss this point, 
together with other conclusions, in Sect.~5. 

Recently Tkachov has also discussed the possibility of 
performing expansions near threshold \cite{TKA}. However, his 
prescription has been applied in \cite{TKA} only to the 
discontinuity of the 1-loop 2-point diagram with two masses. The 
non-trivial interplay of several small scales that characterize 
the threshold region is seen only in more complicated loop 
integrals.\\

{\large\bf \noindent 2. Heuristic motivation and examples}\\

\noindent A typical explicit 
formula for the asymptotic expansion of a given
Feynman integral $F_{\Gm}$ (corresponding to a graph $\Gm$)
in a given limit looks like
\be
F_{\Gm}
\sim 
\sum_{\gamma}  {\cal M}_{\gm} F_{\Gm},
\label{AE}
\ee
where the sum extends over a certain subset of subgraphs
of the graph, and the operators  ${\cal M}_{\gm}$  perform
Taylor expansions in the variables that are small in $\gamma$. To 
arrive at this result, one can use the following heuristic procedure: 
1. Determine the large and small scales in the problem. 
2. Introduce factorization scales $\mu_i$ and divide the loop integration 
domain into regions in which each loop momentum is considered to be 
of the order of one of the scales in the problem. 3. Perform, in every 
given region, a Taylor expansion in the parameters, which are small in the 
given region. 4. After expansion, ignore all factorization scales and 
integrate over the entire loop integration domain in every region.

One can easily reproduce the general formulae and 
combinatorical structure of the large-mass and large-momentum 
expansion in this way. The non-trivial point to justify is 4., which 
also guarantees the homogeneity of the expansion formula. In order 
for 4. to be valid it is essential to use dimensional (or analytic) 
regularization for the Feynman integral $F_\Gamma$, even if $F_\Gamma$ is 
finite in four dimensions. Loosely speaking, 4. 
follows in dimensional regularization from the property that all 
integrals without scale vanish. For off-shell limits of Feynman diagrams, 
the above procedure can indeed be justified \cite{Ch} in terms of the 
$R^*$ \cite{Rstar} and $R^{-1}$-operations. In this section we use these 
heuristic rules to treat two one-loop and one two-loop example and 
demonstrate that in each case the result agrees with the expansion of the 
exact result.\\

\begin{figure}[t]
   \vspace{-4.5cm}
   \epsfysize=30cm
   \epsfxsize=20cm
   \centerline{\epsffile{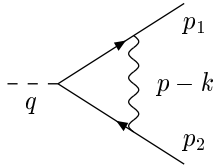}}
   \vspace*{-21.5cm}
\caption[dummy]{\small One-loop vertex integral. Solid (wavy) lines 
denote massive (massless) propagators. \label{fig1}}
\end{figure}
{\em Example 1.}
As our first example we consider a one-loop vertex integral with 
two equal and one zero mass, which appears for example in the 
form factor for $\gamma^*\to \bar{Q}(p_1) Q(p_2)$ with quarks 
$Q$ of mass $m$, see Fig.~\ref{fig1}. In the following we will 
also refer to massive lines as quarks and massless lines as gluons 
in general. We take $p_1^2=p_2^2=m^2$ and 
define $q=p_1+p_2$ and the relative momentum $p=(p_1-p_2)/2$. The 
threshold region is characterized by 
\be
y\equiv m^2-\frac{q^2}{4} = p^2 \ll q^2.
\ee
We also define $\hat{y}=y/q^2$ as the dimensionless parameter of the 
threshold expansion. The parameter $\hat{y}$ is related to a
more standard parameter 
$\beta=(1-4 m^2/q^2)^{1/2}$ through $\beta=\sqrt{-4\hat{y}}$. Above 
threshold $\hat{y}<0$. 
It is useful to choose a frame in which 
$q=(q_0,\vec{0})$, $p_1=(p_0,\vec{p})$, $p_2=(p_0,-\vec{p})$ and 
$p=(0,\vec{p})$. In this frame, the two massive particles move 
slowly and can be considered as non-relativistic. The following three 
scales are relevant to the threshold kinematics: 
$q\equiv \sqrt{q^2}\sim 2 m$, the centre-of-mass energy, 
$\sqrt{-y}=|\vec{p}\,|$, the relative momentum of the 
quarks and $y/q\sim (p_{1,2})_0-m$, the (non-relativistic) energy of the 
quarks. Accordingly, the loop momentum can be either
\begin{eqnarray}
\label{terminology}
\mbox{{\em hard} (h):} && k_0\sim q,\,\,\vec{k}\sim q, \nonumber\\
\mbox{or {\em small} (sm):} &&  k_0\,\simlt \,\sqrt{y},\,\,
\vec{k}\,\simlt\, \sqrt{y}.
\end{eqnarray}
Because there are two small scales, when the loop momentum is small, 
it can be either
\begin{eqnarray}
\label{softterm}
\mbox{{\em potential} (p):} && k_0\sim y/q,\,\,\vec{k}\sim\sqrt{y}, 
\nonumber\\
\mbox{{\em soft} (s):} && k_0\sim \sqrt{y},\,\,\vec{k}\sim\sqrt{y},\\
\mbox{or {\em ultrasoft} (us):} &&  k_0\sim y/q,\,\,\vec{k}\sim y/q.
\nonumber
\end{eqnarray}
The terminology implies a `canonical' routing of the large external 
momentum $q$ through the massive lines of the graph only. For example, 
in Fig.~\ref{fig1}, one should assign the momentum $k\pm q/2$ to
the massive lines. Other routings are possible, but scaling arguments 
become less transparent then. The distinction between potential, soft 
and ultrasoft loop momentum is made only after, in the small 
momentum region, the integration over 
the zero-component of the loop momentum has been carried out by picking 
up the residues of the poles in $k_0$. The potential region is 
then associated with quark propagator poles and $k_0\ll |\vec{k}|$, 
because the massive particles are non-relativistic. The soft and 
ultrasoft regions arise from the gluon propagator poles. We shall see 
that in general (but not in this example) 
the characteristic momentum of on-shell massless particles in the 
small momentum region can be either $\sqrt{y}\sim m v$ or 
$y/q\sim m v^2$, where 
$v$ is the relative velocity of the two external quarks. (When 
the momentum of a massless line is in the potential region, the massless 
particle is off-shell by an amount of order $y$.)
With these preliminaries the scalar integral of Fig.~\ref{fig1} 
is given by
\be
I_1\equiv \int \!\frac{[dk]}{(k^2+q\cdot k-y) (k^2-q\cdot k-y) 
(k-p)^2},
\label{ex1}
\ee
using the kinematic variables introduced above. The standard 
$+i0$-prescriptions are implicitly understood in the propagators and the 
integration measure is defined by $[dk]\equiv e^{\epsilon\gamma_E}\,
d^dk/(i\pi^{d/2})$ with $d=4-2\epsilon$ and $\gamma_E=0.577216\ldots$. 
The renormalization scale of dimensional regularization is set to 1.

When the loop momentum $k$ is hard, the integrand is expanded in 
$y$ and $p$. (Because $p\cdot q=0$, terms odd in $p$ vanish and the 
expansion produces only powers of $\hat{y}$.) The leading term is
\be
I_1^h= \int \!\frac{[dk]}{k^2 (k^2+q\cdot k) (k^2-q\cdot k)} = 
e^{\epsilon\gamma_E}\left(\frac{4}{q^2}\right)^{\!1+\epsilon}\!
\left(\!-\frac{1}{2}\right) \frac{\Gamma(\epsilon)}{1+2\epsilon}.
\ee
Higher-order terms in $y$ are calculated easily. The integral is evaluated 
most directly by reducing the number of propagators to two through 
partial fractioning. If one does not use partial fractions, the 
calculation of the integral with Feynman parameters leads to 
singularities inside the Feynman parameter integration domain when 
$d=4$, which indicate the presence of the Coulomb singularity in 
$I_1$ as $y\to 0$. This singularity is regulated by dimensional 
regularization as conventional ultraviolet and infrared singularities 
are, and $I_1^h$ is well-defined.

The contribution from the hard region corresponds to the `naive' Taylor 
expansion of the integrand in $y$ and $p$. However, the integral receives 
an important (in fact dominant) contribution 
from the small loop-momentum region, 
where (as we will verify shortly) the spatial loop momentum is of order 
$\sqrt{y}$. Inspecting the quark 
propagator, we see that it can now be expanded in $k_0^2$:
\be
\label{softapprox}
\frac{1}{k_0^2-\vec{k}^2+q_0 k_0-y} = 
\sum_{n=0}^\infty\frac{(-k_0^2)^{n}}{(-\vec{k}^2+q_0 k_0-y)^{n+1}}.
\ee
The leading contribution to $I_1$ from the small momentum region is then
\be
\label{small}
I_1^{sm} = \int \!\frac{[dk]}{(-\vec{k}^2+q_0 k_0-y) (-\vec{k}^2-q_0 k_0-y) 
(k_0^2-(\vec{k}-\vec{p}\,)^2)}.
\ee
We now perform the integration over $k_0$ by closing the contour in the 
upper complex $k_0$-plane. 

Let us consider first the contribution 
from the pole of the massive (`quark') propagator at 
$k_0=-(\vec{k}^2+y)/q_0+
i0$. Since $|\vec{k}|\ll q$ by assumption, we have $k_0\ll |\vec{k}|$. 
As a consequence we might have expanded the massless propagator in 
(\ref{small}) in $k_0^2$ before picking up the residue from this pole. 
After $k_0$-integration one obtains 
a two-point function with one mass $y$ in $d-1$ dimensions with the 
result
\be
\label{ex1pot}
I_1^p = \frac{(-1)}{q}\,e^{\epsilon\gamma_E}
\int\!\frac{d^{d-1}\vec{k}}{\pi^{d/2-1}}\,
\frac{1}{(\vec{k}^2+y) (\vec{k}-\vec{p}\,)^2} = 
e^{\epsilon\gamma_E}\,
\frac{y^{-\epsilon}}{\sqrt{q^2 y}}\,\frac{\sqrt{\pi}\,\Gamma(\epsilon+1/2)}
{2\epsilon}.
\ee
Because the integrand contains only one scale, the loop momentum is 
dominated by $\vec{k}\sim\sqrt{y}$. This justifies the scaling given 
in (\ref{terminology}) for potential loop momenta.
Note that higher-order terms in the expansion in $k_0^2$ yield zero, 
because after integration over $k_0$ positive powers of $k_0$ result in 
massless tadpole integrals. Note also that we could have obtained 
(\ref{ex1pot}) by expanding all terms that are small in the 
potential region before integration over $k_0$. 
The contribution to the threshold expansion from the potential 
region can then easily be estimated, 
because by construction all terms in the denominator have the 
same scaling in $y$. 
Taking into account that $[dk]\sim y^{5/2}/q$ for 
potential $k$, we obtain $I_1^{p}\sim 1/\sqrt{q^2 y}$ as expected for the 
Coulomb singularity and born out by (\ref{ex1pot}). 

Now consider the contribution from the pole of the massless propagator 
in (\ref{small}) located at $k_0=-|\vec{k}|+i0$. (We have shifted 
$k\to k+p$. If one does not perform this shift, the combination 
$k+p$ should be considered ultrasoft, which implies a cancellation 
between $k$ and $p$, so that the scaling rules can not be applied 
to $k$ in a straightforward way.) The result is
\be
\label{gluonpole}
I_1^{s/us} = e^{\epsilon\gamma_E}\int\!\frac{d^{d-1}k}{\pi^{d/1-2}}
\frac{1}{|\vec{k}| [q_0 |\vec{k}|+(\vec{k}^2+2\vec{p}\cdot\vec{k})] 
[q_0 |\vec{k}|-(\vec{k}^2+2\vec{p}\cdot\vec{k})]}.
\ee
Expanding in the small terms $\vec{k}^2+2\vec{p}\cdot\vec{k}$, the 
integral becomes a tadpole and is zero to all orders in the expansion. 
Hence, there is no contribution from the gluon pole. At first sight 
this looks incorrect, 
because the integral (\ref{gluonpole}) is clearly non-zero. However, the 
non-zero contribution can come only from the region where 
$\vec{k}^2+2\vec{p}\cdot\vec{k}$ is comparable to $q_0|\vec{k}|$, which 
requires $k\sim q$. This contribution is already included in the hard 
contribution above and (\ref{gluonpole}) {\em must} be expanded to 
avoid double-counting. Because the integral $I_1^{s/us}$ vanishes, we 
can not decide whether 
the gluon was soft or ultrasoft. 

There is a useful short-cut to arrive at this result. We can perform the 
approximations appropriate to the soft or ultrasoft region before 
integrating over $k_0$ to obtain, to leading order in this region, 
\be
\label{ex1sus}
I_1^{s/us} = \int\!\frac{[dk]}{[q_0 k_0+i0] 
[-q_0 k_0+i0] \,k^2}.
\ee
This integral is ill-defined, because the poles at $k_0=0$ pinch the 
integration contour. However, since the quark propagator poles have 
already been taken into account through the potential region, the 
previous integral should be understood as the contribution from the 
gluon pole only. Then one immediately arrives at a vanishing tadpole 
integral. Since by expanding all small quantities in the denominator, 
all remaining terms have the same scaling in $\sqrt{y}$, we have 
homogeneity (manifest power counting) also for the soft and ultrasoft 
regions.
  
One can now verify that the threshold 
expansion reproduces the expansion of the exact one-loop result: 
\bea
I_1 &=& e^{\epsilon\gamma_E} \,y^{-1-\epsilon} \,
{}_2F_1\left(\frac{1}{2},
1+\epsilon,\frac{3}{2};-\frac{1}{4\hat{y}}\right) \nonumber\\
&=& e^{\epsilon\gamma_E}\left(\frac{4}{q^2}\right)^{1+\epsilon}\!
\Bigg\{\frac{(4\hat{y})^{-\epsilon}}{\sqrt{\hat{y}}}\,
\frac{\sqrt{\pi}\Gamma(\epsilon+1/2)}{8\epsilon} \\
&&\, -\frac{\Gamma(\epsilon)}{2 (1+2\epsilon)}\sum_{n=0}^\infty 
\frac{\Gamma(1+\epsilon+n)}{\Gamma(1+\epsilon)}
\frac{1+2\epsilon}{1+2\epsilon+2 n}\,\frac{(-4\hat{y})^n}{n!}
\Bigg\}. \nonumber
\eea
Note that in this 
example, contrary to those we discuss later, the Taylor expansions 
of the integrands do not generate additional poles in $\epsilon$. The poles 
in $\epsilon$ in both the hard and the potential 
contribution come from the infrared pole which is present in the 
original integral $I_1$.\\

\begin{figure}[t]
   \vspace{-3.5cm}
   \epsfysize=25.2cm
   \epsfxsize=16.8cm
   \centerline{\epsffile{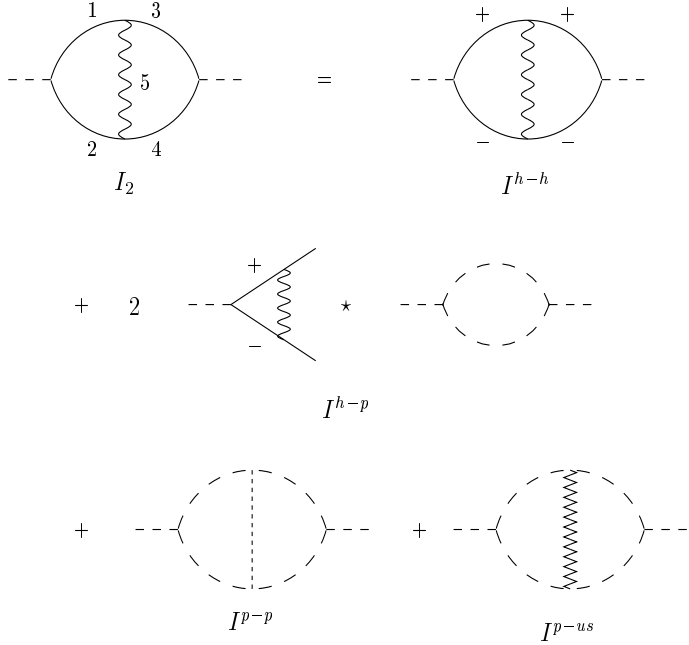}}
   \vspace*{-11.5cm}
\caption[dummy]{\small Diagrammatic representation of the threshold 
expansion for $I_2$. Solid lines on the left-hand side denote massive 
propagators, wavy lines denote massless propagators. On the right-hand side 
a solid line with `$\pm'$ denotes (hard) `on-shell' propagators of 
form $1/(k^2\pm q.k)$ for line momentum $k\pm q/2$, a wavy line 
stands for a hard massless line. The dashed line denotes a 
potential massive line and the dotted line a potential massless line. 
The zigzagged line is an ultrasoft massless line. \label{fig2}}
\end{figure}
{\em Example 2.}
We now consider the two-loop 2-point master integral. This is the first 
case, relevant to the two-loop corrections to heavy-quark currents, 
where the hard, potential and ultrasoft 
regions contribute in a non-trivial way to the 
threshold expansion. The exact result for arbitrary $\hat{y}$ serves as 
a check of the expansion method. The diagram, shown in Fig.~\ref{fig2}, 
is given by
\be
I_2\equiv\int \!\frac{[dk] [dl]}{(k^2+q\cdot k-y) (k^2-q\cdot k-y)
(l^2+q\cdot l-y) (l^2-q\cdot l-y) (k-l)^2 } 
\label{mast1}
\ee
and the lines in Fig.~\ref{fig2} are enumerated according to the 
order in which they appear in the denominator in (\ref{mast1}). 
Expansions of this diagram in various limits of momenta and masses
have been considered in \cite{Davy}. 
The `canonical' routing of the external momentum for this diagram assigns 
$\pm q/2$ to each massive line. With this assignment, the loop momenta 
satisfy the scaling rules (\ref{terminology}) and (\ref{softterm}) above. 
For the threshold expansion of any two-loop diagram, one considers 
the following nine loop momentum regions: h-h: both loop momenta are hard; 
h-p (s, us): one loop momentum hard, the other potential (soft, ultrasoft); 
p-p: both loop momenta potential; p-s (us): one loop momentum potential, 
the other soft (ultrasoft); s-s: both loop momenta soft; s-us: one 
loop momenta soft, the other ultrasoft; us-us: both loop momenta ultrasoft. 
In general, to obtain the contributions 
from all possible subgraphs it is necessary to consider different 
assignments of loop momenta than the one chosen in (\ref{mast1}). As 
we discuss shortly, some of these regions do not contribute to the case at 
hand. The structure of the non-vanishing terms in the expansion 
is shown in Fig.~2. Note that $I_2$ is finite in four dimensions. 

When both loop momenta are of order $q$, we can expand the integrand 
in $y$. After partial fractioning, the resulting integrals are of type
\be
\label{Jints}
J_\pm(a_1,\ldots,a_5)\equiv \int\!\frac{[dk] [dl]}{[-k^2]^{a_1} 
[-l^2]^{a_2} [-(k-l)^2]^{a_3} [-(k^2+q\cdot k)]^{a_4} 
[-(l^2\pm q\cdot l)]^{a_5}}. 
\ee
The integrals of type $J_+$ can be reduced to Gamma-functions 
\cite{GRA} through 
recurrence relations derived from integration by parts \cite{IBP}. 
The integrals of type $J_-$ can be expressed in Gamma-functions and 
the integrals $J_-(0,0,a_3,a_4,a_5)$. These integrals can be reduced 
to $J_-(0,0,1,1,1)$, using a simplified version of the results of 
\cite{TAR}. Then $J_-(0,0,1,1,1)$ (or a more convenient input integral) 
is calculated explicitly by Feynman parameters in an 
expansion in $\epsilon$. We obtain
\be 
\label{h-h}
q^2 I_2^{h-h} = \pi^2\left(\frac{1}{\epsilon}-2\ln q^2+6\ln 2\right) + 
 21\zeta(3) - 4 (8+3\pi^2)\,\hat{y} + {\cal O}(\hat{y}^2).
\ee

The h-h region is already the most difficult one. When one loop momentum 
is hard, and the other small, 
one of the three subgraphs $\gamma_1=\{1,2,5\}$, 
$\gamma_2=\{3,4,5\}$ and $\gamma_3=\{1,2,3,4\}$ can be hard. (The 
numbers refer to those in Fig.~2.) The lines 
in the hard subgraph $\gamma_i$ are expanded in $y$ and $l$ (if 
$k$ is hard) and the quark propagators 
not in $\gamma_i$ are expanded 
in the zero-components squared of momenta as in 
(\ref{softapprox}). 
The resulting contribution has the factorized form
\be
\label{fact}
F_{\Gamma/\gamma_i} \circ {\cal T}_y F_{\gamma_i},
\ee
i.e. the one-loop integral ${\cal T}_y F_{\gamma_i}$, 
obtained by Taylor expansion of $F_{\gamma_i}$ in $y$ and $l$, generates a 
local vertex, which is inserted into the reduced diagram $\Gamma/\gamma_i$ 
obtained from $\Gamma$ by shrinking all lines of $\gamma_i$ to a 
point and expansion of the remaining massive 
propagators in the zero-components squared of the momenta. 

For our example, one obtains a vanishing integral for $\Gamma/\gamma_3$ 
and equal contributions from $\gamma_1$ and $\gamma_2$. In the latter 
case, the non-vanishing contribution arises if the loop momentum 
of $\Gamma/\gamma_{1,2}$ is potential, but one obtains zero if it is 
soft or ultrasoft. The leading 
contribution from the h-p region is then (see Fig.~2)
\be 
\label{h-s1}
I_2^{h-p} = 2\int\!\frac{[dl]}{(-\vec{l}^2+q_0 l_0-y) 
(-\vec{l}^2-q_0 l_0-y)}\,\int\!\frac{[dk]}{k^2 (k^2+q\cdot k) 
(k^2-q\cdot k)},
\ee
where we expanded all terms that are small in the h-p region. 
It follows from power-counting that this integral contributes at 
order $\sqrt{\hat{y}}$. Keeping also the next term in the h-p region, 
we obtain
\be 
\label{h-p}
q^2 I_2^{h-p} = 8\pi\left(\frac{1}{\epsilon}-\ln q^2-\ln y\right) 
\sqrt{\hat{y}} -\frac{32 \pi}{3}\left(\frac{1}{\epsilon}-\ln q^2-\ln y+
\frac{7}{3}\right) \hat{y}^{3/2} + {\cal O}(\hat{y}^{5/2}).
\ee
When $l$ is soft or ultrasoft, the integrand is 
almost the same as for (\ref{h-s1}), except that $\vec{l}^2$ is 
expanded for ultrasoft $l$ and $\vec{l}^2+y$ is expanded for soft
$l$. In both cases, the resulting integrals vanish as already mentioned.

When both loop momenta are small, we can use (\ref{softapprox}) on all 
massive propagators and then take the poles in $k_0$ and $l_0$. The 
resulting three-dimensional integrals are then divided into regions 
according to (\ref{softterm}). As illustrated in example 1, one can 
also perform the appropriate approximations before integration over 
the zero components of loop momenta provided certain pinching quark poles 
in the soft or ultrasoft region are ignored as above.

When both loop momenta are potential, all propagators are expanded in their 
zero-components squared. After picking up the residues from the 
remaining quark poles in $k_0$ and $l_0$, the resulting $d-1$-dimensional 
2-loop integral is
\bea
I_2^{p-p} &=& -\frac{1}{q^2}\,e^{2\epsilon\gamma_E}
\int\!\frac{d^{d-1}\vec{k}}{\pi^{d/2-1}}\frac{d^{d-1}\vec{l}}{\pi^{d/2-1}}\,
\frac{1}{(\vec{k}^2+y)(\vec{l}^2+y) (\vec{k}-\vec{l}\,)^2}\nonumber\\
&=& \frac{y^{-2\epsilon}}{q^2}\,e^{2\epsilon\gamma_E} \,
\frac{\pi\Gamma(\epsilon+1/2)\Gamma(\epsilon-1/2)}{2\epsilon}
\eea
as shown in Fig.~2. The next term and all higher order terms 
in the expansion in the p-p region 
vanish for the same reason as in example 1, hence
\be
\label{s-s}
q^2 I_2^{p-p} =
\pi^2\left(-\frac{1}{\epsilon}-2+4\ln 2+2\ln y\right).
\ee

We next consider the potential-ultrasoft region. 
Again we should consider all three cases, where
one of the subgraphs $\gamma_i$ is potential and the remaining lines 
are ultrasoft. The only case that does not lead to vanishing scaleless 
integrals is potential $\gamma_3$, in which case the gluon line is 
ultrasoft. With $k$ potential and $l$ ultrasoft, the leading term is 
\bea 
I_2^{p-us} &=& \int\!\frac{[dk] [dl]}{(-\vec{k}^2+q_0(k_0-l_0/2)-y)  
(-\vec{k}^2-q_0(k_0-l_0/2)-y)} \nonumber\\
&&
\hspace*{-1cm}\cdot\,\frac{1}{(-\vec{k}^2+q_0(k_0+l_0/2)-y) 
(-\vec{k}^2-q_0(k_0+l_0/2)-y) (l_0^2-\vec{l}^2)}
\eea
where the last (massless) propagator is now not expanded in $l_0$. 
(We have chosen 
a more symmetric loop momentum routing than in (\ref{mast1}).) Since 
$[dk]\sim y^{5/2}/q$ and $[dl]\sim y^4/q^4$, power-counting  
tells us that this contribution scales as $\sqrt{y}$. The integral is 
calculated by examining the poles in $k_0$ and $l_0$ and closing the 
integration contours in $k_0$, $l_0$ such that the number of terms is 
minimized. One then arrives at
\be
I_2^{p-us} = \frac{2}{q^2}\,e^{2\epsilon\gamma_E}
\int\!\frac{d^{d-1}\vec{k}}{\pi^{d/2-1}}\frac{d^{d-1}\vec{l}}{\pi^{d/2-1}}\,
\frac{1}{(\vec{k}^2+y)\,\,\vec{l}^2\,(2 (\vec{k}^2+y)+q_0 |\vec{l}|)}.
\ee
Note that although $|\vec{k}|\gg|\vec{l}|$ in the region under 
consideration, the integral does not factorize in a manner 
comparable to (\ref{fact}), because of the presence of $q_0\gg |\vec{k}|$ 
in the integrand. This can be phrased as the statement that 
integrating out the soft momenta does not yield a local 
interaction vertex with respect to the ultrasoft scale. This is in 
contrast to integrating out the hard momentum, which does result in a 
local interaction with respect to the small scales, see (\ref{fact}),  
where ${\cal T}_y F_{\gamma_i}$ is polynomial in its (small) external 
momenta. Despite this fact, the above integral is 
effectively a succession of one-loop integrals. The result, keeping 
again the next term in the expansion in this region, reads
\be
\label{s-us}
q^2 I_2^{p-us} = 8\pi\left(-\frac{1}{\epsilon}-8+10\ln 2-\ln q^2+
3 \ln y\right) \left(\sqrt{\hat{y}}-\frac{4\hat{y}^{3/2}}{3}\right) 
+ {\cal O}(\hat{y}^{5/2}).
\ee
A technical comment on the calculation of higher-order terms 
in $\hat{y}$ in the potential-ultrasoft region is in order. Expansion 
in the zero-components of the line momenta squared results in many terms, 
for each of which separately the integral over the semi-circle at infinity 
does not vanish upon closing the contours in the complex plane. As a 
consequence one has to take care of closing the contour in the same 
half plane for all terms, since the contribution from the semi-circle at 
infinity adds to zero only in the sum.

Finally, when one momentum is potential and the other soft, we find 
scaleless integrals. Likewise when both momenta are either soft or 
ultrasoft, one obtains only vanishing integrals. This concludes the 
list of all relevant momentum regions. 

In general, the h-h and p-p region contribute only to even powers of 
$\sqrt{\hat{y}}$ and the h-p and p-us region contribute only to odd powers. 
The contribution from each region separately contains poles as $\epsilon 
\to 0$. The poles cancel between the h-h and p-p region and the 
h-p and p-us region, leaving logarithms of $\hat{y}$. The terms 
computed above combine to the finite threshold expansion
\bea
\label{resi2}
q^2 I_2 &=& 2\pi^2\ln(32\hat{y}) + 21\zeta(3) + 
16\pi\Big(\ln(32\hat{y})-4\Big)\sqrt{\hat{y}}
-4\,(8+3\pi^2) \,\hat{y}  \nonumber\\
&&\,-\frac{64\pi}{3}\left(\ln(32\hat{y})-
\frac{17}{6}\right)\hat{y}^{3/2} + \ldots.
\eea
The integral $I_2$ is known exactly for any $\hat{y}$. Using 
the result of \cite{I2}, we have
\be
q^2 I_2 = F(1)+F(z^2)-2 F(z),
\ee
where 
\bea
F(z) &=& 6\,\mbox{Li}_3(z)-
4 \ln z \,\mbox{Li}_2(z)-\ln^2(z)\,\ln(1-z), \\
z &=& -\frac{1-i\sqrt{4\hat{y}}}{1+i\sqrt{4\hat{y}}}.
\eea
Taking care of correctly continuing the logarithms in $F(z^2)$ to the 
second sheet when $\hat{y}<1/4$, the expansion of the exact result 
reproduces (\ref{resi2}). We have verified that this agreement 
persists to higher orders in $\hat{y}$.\\
  
\begin{figure}[t]
   \vspace{-4.5cm}
   \epsfysize=30cm
   \epsfxsize=20cm
   \centerline{\epsffile{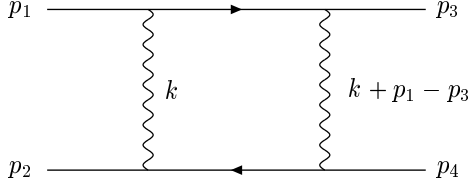}}
   \vspace*{-21.2cm}
\caption[dummy]{\small The box graph. Solid lines massive 
propagators, wavy lines massless. \label{fig3}}
\end{figure}
{\em Example 3.} In the previous two examples, soft (on-shell) massless 
lines (gluons) did not contribute to the expansion. To obtain a 
non-vanishing result from this region, at least two gluons have 
to be exchanged between the massive lines (quarks). Consider 
a one-loop contribution to $2\to2$ 
scattering at small relative momentum, Fig.~\ref{fig3}, and define 
$p_{1/2}=q/2\pm p$, $p_{3/4}=q/2\pm p'$, where $q+p_1+p_2=p_3+p_4$, 
$p=(p_1-p_2)/2$ and $p'=(p_3-p_4)/2$. All external lines are on-shell, 
$p_i^2=m^2$, and with $y=m^2-q^2/4$ as before, we have $p^2=p'^2=y$. 
We also introduce $t=(p'-p)^2$ and choose a frame where 
$q=(q_0,\vec{0})$, $p=(0,\vec{p}\,)$ and $p'=(0,\vec{p}^{\,\prime})$. 
Then the threshold region is characterized by
\begin{equation}
t\sim y\ll q^2.
\end{equation}
The integral represented by Fig.~\ref{fig3} is
\begin{equation}
I_3 \equiv \int \!\frac{[dk]}{((k+p)^2+q\cdot k-y) ((k+p)^2-q\cdot k-y) 
k^2 (k+p-p')^2}.
\label{ex3}
\end{equation}
The threshold expansion is obtained by letting the loop momentum be 
hard, potential, soft or ultrasoft and by expanding the integrand in 
all quantities that are small in the given region. In the following, 
we collect all terms up to order $y^0/q^4$. (Recall that the tree diagram 
for $2\to 2$ scattering is $1/t \sim 1/y$.) 

When the loop momentum is hard, the integrand is expanded in 
$y$, $p$ and $p'$ and one obtains
\be
(q^2)^{2+\epsilon} I_3^h = 
(q^2)^{2+\epsilon} 
\int \!\frac{[dk]}{(k^2)^2 (k^2+q\cdot k) (k^2-q\cdot k)} = 
-\frac{8}{3}.
\ee

When the loop momentum is potential, all four propagators are expanded 
in $k_0^2$. This corresponds to picking up the quark propagator pole 
in the integration over $k_0$. Since the leading contribution scales 
as $y^{-3/2}$, subleading terms have to be included to achieve 
${\cal O}(y^0)$ accuracy in the threshold expansion. Defining $\hat{t}=
t/q^2$, the result is 
\be
(q^2)^{2+\epsilon} I_3^p = \frac{\pi}{\hat{t}\sqrt{\hat{y}}} 
\left(\frac{1}{\epsilon}-\ln(-\hat{t})\right)
+ {\cal O}(\hat{y}^{1/2},\hat{t}^{1/2}).
\ee
The subleading term actually vanishes in four dimensions.

Let the loop momentum be ultrasoft. It is not possible to route the 
external momentum $p$ or $p'$ such that both gluon lines are ultrasoft 
simultaneously. Thus we get 
\begin{equation}
I_3^{us} =\frac{1}{t} \int \!\frac{[dk]}{[q_0 k_0+i0] [-q_0 k_0+i0]  
k^2} = 0,
\end{equation}
where the pinch at $k_0=0$ must be ignored, because it is taken into 
account by the potential region.

However, there arises a non-zero contribution from the gluon poles, when 
the loop momentum is soft. In this case, to leading order in the soft 
approximation, the integral simplifies to 
\begin{equation}
I_3^s  = \frac{1}{q^2} 
\int \!\frac{[dk]}{k_0^2 k^2 (k+p-p')^2}.
\end{equation}
Closing the $k_0$-contour, we obtain
\begin{equation}
I_3^s = \frac{1}{q^2} 
\int \!\frac{d^{d-1}\vec{k}}{\pi^{d/2-1}}\, 
\frac{1}{\vec{k}^2-(\vec{k}+\vec{p}-\vec{p}^{\,\prime})^2} 
\left(\frac{1}{|\vec{k}|^3}-\frac{1}{|\vec{k}+\vec{p}-\vec{p}^{\,\prime}|^3}
\right).
\end{equation}
The singularity at $\vec{k}^2=(\vec{k}+\vec{p}-\vec{p}^{\,\prime})^2$ in the 
first factor is cancelled only in the sum of the two 
terms. Thus we cannot shift 
the momentum in the second term alone to make it equal to the first.  
We can calculate both terms separately, if we introduce a $+i0$-prescription 
in the first factor. After this we can shift the loop momentum and obtain
\begin{equation}
I_3^s = \frac{1}{q^2} 
\int \!\frac{d^{d-1}\vec{k}}{\pi^{d/2-1}}\, 
\frac{1}{(\vec{k}^2)^{3/2}} 
\left(\frac{1}{-2\vec{k}\cdot (\vec{p}-\vec{p}^{\,\prime})+t+i0}+
\frac{1}{-2\vec{k}\cdot (\vec{p}-\vec{p}^{\,\prime})+t-i0}\right), 
\end{equation}
Now the integrals can be calculated by standard methods, taking care 
of the different $i0$-prescriptions in the two terms. Again 
keeping the first subleading 
term to the threshold expansion from this region, we find 
\be
(q^2)^{2+\epsilon} I_3^s = \frac{4}{\hat{t}}\left(
-\frac{1}{\epsilon}+\ln(-\hat{t})\right) + 
\frac{16\hat{y}}{3\hat{t}} \left(\frac{1}{\epsilon}-\ln(-\hat{t})
\right) + \frac{8}{3} + {\cal O}(\hat{y},\hat{t}).
\ee
The contribution from the soft region is rather peculiar from the 
conceptual point of view. Both gluons in the box subgraph have energies 
of order $\sqrt{y}$, parametrically larger than the characteristic 
energy scale $y/q$ for the non-relativistic quarks. Hence one can 
imagine the exchange of the two gluons as taking place on a much shorter 
time scale than the one that is characteristic for a non-relativistic 
system. The intermediate quark pair becomes off-shell by an amount 
$\sqrt{q^2 y}$ through the interaction with soft gluons, while the 
off-shellness of the quarks remains of order $y$ in the potential and 
ultrasoft regions. The soft region thus can be considered intermediate 
between the hard region (off-shellness of order $q^2$) and the two 
other small momentum regions. However, while a hard subgraph can be 
interpreted as a local interaction vertex, the soft box subgraph, when 
it appears as a subgraph in a 2-loop diagram gives rise to a 
vertex which is non-local in its external three-momenta but local 
in the zero-components of the external momenta as we discuss in more 
detail in Sects.~3 and 4. 
We also understand why the soft region did not appear in the 
previous two examples. Since an external quark {\em pair} or a quark pair 
that couples to the external current is off-shell by an amount of order 
$y$, after a quark interacts with a soft gluon which puts it 
off-shell by an amount of order $\sqrt{q^2 y}$, it has to interact 
with at least a second soft gluon to reduce its off-shellness to 
$y$ again.

The threshold expansion can be compared with the very 
simple exact (up to terms that vanish as $\eps\to 0$) result for the 
box diagram \cite{BD90}:
\bea
(q^2)^{2+\epsilon} I_3 &=& \frac{2}{\hat{t}\sqrt{\hat{y}}}
\left[\frac{\pi}{2}-\arctan\sqrt{4\hat{y}}
\,\right] \left(\frac{1}{\epsilon}-\ln(-\hat{t})\right)\nonumber\\
&=& \left[\frac{\pi}{\hat{t}\sqrt{\hat{y}}}-
\frac{4}{\hat{t}}+\frac{16\hat{y}}{3\hat{t}}+\ldots\right] 
\left(\frac{1}{\epsilon}-\ln(-\hat{t})\right).
\eea
The expansion in $\hat{y}$ and $\hat{t}$ agrees with the sum of the 
terms computed above. Note the cancellation between the hard region and 
the constant from the soft region, which has to repeat itself 
in higher orders due to the factorized $t$-dependence of the 
exact result.\\

{\large\bf \noindent 3. General structure of the threshold expansion}\\ 

In the previous section we have been rather explicit and 
worked out the threshold expansion on simple examples. In this 
section we abstract from these examples general prescriptions
for the threshold expansion ($y=m^2-q^2/4$ and other small parameters 
such as $\hat{t}$ in example 3 tend to zero at fixed $m^2$) of an arbitrary 
diagram which contains two paths of lines with `slow' particles  
with mass $m$ (`quarks') and some massless particles (`gluons'). 
(Internal loops of massive particles can be included straightforwardly: 
If all momenta that connect to the internal heavy quark loop are 
small, the integrand is expanded in external momenta over the mass 
and reduces to a series of local operators analogous to the 
Euler-Heisenberg effective Lagrangian. If large momentum 
flows through part of the graph, the corresponding lines are 
not expanded. With this in mind, the following discussion is adapted 
to diagrams without internal heavy quark loops.) The two 
paths of massive lines are assigned $q/2$ and $-q/2$ of the large 
external momentum $q$. The two paths can be either disconnected 
as in the box graph example or they can be joined together by 
a hard external vertex that ejects large momentum $q$ into the 
graph, such as in the two-point and three-point functions 
discussed in Sect.~2. We also assume a frame such that 
$q=(q_0,\vec{0})$. The loss of explicit covariance is intrinsic to 
the threshold problem and, while an explicitly covariant formulation 
is possible, it would be exceedingly cumbersome.

We arrive at our general prescription in three steps, which correspond 
to dealing with the hierarchy of scales sequentially. To arrive at the 
first form of the expansion, we consider any loop momentum to 
be either large or small. However, at this step, we do not 
identify small momenta as potential, soft or ultrasoft. In particular, 
when a loop momentum that flows through a quark line is small, 
the zero-component squared of the loop momentum in the quark 
propagator is small, see (\ref{softapprox}). 
Thus, at the first step, we arrive at the following expansion 
\be
\label{exp_sm_la}
F_\Gamma = \sum_{\gm}  {\cal T}_{k_{i0}^2} F_{\Gamma/\gamma} \circ 
{\cal T}_{k^{\gm},y} F_{\gamma},
\ee
where the sum is in one-particle-irreducible (1PI) subgraphs of 
the given graph $\Gm$. The operator
${\cal T}_{k^{\gm}, y}$ in the second factor performs the Taylor expansion
of the Feynman integral $F_{\gamma}$ in $y$ and the loop momenta 
$k^\gamma$ 
of $\Gm$ which are external with respect to $\gm$. (These are 
the loop momenta of the reduced graph $\Gamma/\gamma$, which is 
obtained by shrinking all lines of $\gamma$ to a point.) 
The operator ${\cal T}_{k_{i0}^2}$ in the first factor 
performs the Taylor expansion
of the Feynman integral $F_{\Gamma/\gm}$ associated with the reduced
graph in the squares of the zero-components of the loop momenta that
flow through the quark lines of the given paths. The above 
equation has the combinatorical structure of a factorized expression. 
It is easy to see that (\ref{exp_sm_la}) is equivalent to 
constructing a non-relativistic effective Lagrangian \cite{CL}. 
The factors ${\cal T}_{k^{\gm},y} F_{\gamma}$ can be associated 
with the insertion of local operators into the diagram and they 
account for all hard loop momenta. The factors 
${\cal T}_{k_{i0}^2} F_{\Gamma/\gamma}$ can be associated with 
diagrams in the effective theory and the hard scale $q$ is not 
present as a scale for loop momenta in this part any more. 
Note that the asymptotic expansions 
for certain limits of on-shell integrals considered in 
\cite{acvs} can be recast into a form similar to (\ref{exp_sm_la}), 
with expansion in $k^2$ rather than $k_0^2$ in the small momentum 
part.

However, the expansion (\ref{exp_sm_la}) is not homogeneous in the 
expansion parameter. The non-homogeneity arises from the small 
momentum parts ${\cal T}_{k_{i0}^2} F_{\Gamma/\gamma}$, which are 
still non-trivial series in the threshold expansion parameter(s).  
To obtain the expansion in a manifestly homogeneous form we have 
to specify further the scale of the loop momenta of the reduced graphs
$\Gamma/\gamma$ in (\ref{exp_sm_la}). At this point we consider 
potential, soft and ultrasoft loop momenta with the scaling rules 
given in (\ref{softterm}). Let $\hat{\Gm}$ be a reduced graph that 
appears in (\ref{exp_sm_la}). Since we have already disposed of all 
hard momentum regions, the `next hardest' region is the region 
of soft loop momenta. Consider a soft subgraph $\gm_s$ of $\hat{\Gm}$, 
i.e. a 1PI subgraph of $\hat{\Gm}$ in which all loop momenta 
$l^{\gm_s}$ scale as $l^{\gm_s}_0\sim\vec{l}^{\gm_s}\sim\sqrt{y}$ and 
the external momenta $k^{\gm_s}$ of $\gm_s$ are either potential or 
ultrasoft. Then $\gm_s$ is expanded in all external 
momenta, which are ultrasoft, and reduces to a local interaction 
with respect to the ultrasoft scale. For a potential external 
momentum, we expand only in its zero component, because the 
spatial component of a potential momenta is of the same order as 
the spatial component of a soft momentum. As a consequence, the 
soft subgraph is a temporally local vertex with respect to the 
potential scale, but it is spatially non-local, because the 
dependence on the three-momenta of the potential external momenta 
is not polynomial. Therefore, defining 
$\hat{F}_{\hat{\Gm}}={\cal T}_{k_{i0}^2} F_{\hat{\Gamma}}$ as 
it appears in (\ref{exp_sm_la}), we arrive at
\be
\label{exp_s_pus}
\hat{F}_{\hat{\Gamma}} = 
\sum_{\gm_s} \hat{F}_{\hat{\Gamma}/\gm_s} \circ 
{\cal T}_{k^{\gm_s}_{us},k^{\gm_s}_{p,0}} \hat{F}_{\gamma_s},
\ee
where the sum is again in all 1PI subgraphs. In addition to the 
Taylor expansions indicated explicitly, the massive propagators 
in $\gm_s$ should be expanded according to
\be
\frac{1}{-(\sum_i\vec{l}^{\gm_s}+\sum_j\vec{k}^{\gm_s})^2\pm 
q_0 (\sum_i\vec{l}^{\gm_s}_0+\sum_j\vec{k}^{\gm_s}_0) - y} 
= \frac{1}{\pm q_0\sum_i\vec{l}^{\,\gm_s}_0} + \ldots
\ee
and pinched poles are simply to be ignored, see the discussion in 
Sect.~2. Notice that these are 
static propagators rather than non-relativistic propagators. 
After inserting (\ref{exp_s_pus}) in (\ref{exp_sm_la}) we have 
achieved factorization of both hard and soft momenta. The 
factors ${\cal T}_{k^{\gm_s}_{us},k^{\gm_s}_{p,0}} \hat{F}_{\gamma_s}$ 
can be interpreted as an instantaneous interaction vertex, 
and thus as contribution to the heavy quark potential. On the other 
hand, since ${\cal T}_{k^{\gm_s}_{us},k^{\gm_s}_{p,0}} \hat{F}_{\gamma_s}$ 
is spatially non-local, the graph 
$\hat{\Gm}/\gm_s$ is interpreted as deleting all lines 
of $\gm_s$ in $\hat{\Gm}$ and replacing them by 
${\cal T}_{k^{\gm_s}_{us},k^{\gm_s}_{p,0}} \hat{F}_{\gamma_s}$. Let us 
give an example. The soft part of the box graph of Sect.~2 
has four potential external quark lines and we have
\be
{\cal T}_{k^{\gm_s}_{us},k^{\gm_s}_{p,0}} \hat{F}_{box} 
\to \frac{c}{q_0^2}\,\frac{1}{[-(\vec{p}-\vec{p}^{\,\prime})^2]^
{1+\epsilon}}
\ee
with a numerical constant $c$. In coordinate space, this gives 
rise to a non-local four-fermion operator. Due to the structure of 
(\ref{exp_s_pus}), the factorization of soft regions can also be 
implemented as an effective Lagrangian, although it is non-local 
in space. In the planar two-loop three-point integral which we 
treat in Sect.~4, we will see how such a non-local operator 
inserted into a reduced diagram contributes to the threshold expansion 
of two-loop diagrams. In Sect.~4 we will also meet another 
non-local vertex for the interaction of two potential quark lines 
with one potential gluon line. One can use power counting for the 
soft region to determine all non-local operators which contribute 
to a given order in the threshold expansion.

The expansion obtained from combining (\ref{exp_sm_la}) and 
(\ref{exp_s_pus}) is still not homogeneous in the threshold expansion 
parameter. The inhomogeneity arises from $\hat{F}_{(\Gm/\gm)/\gm_s}$, 
where hard and soft lines are deleted, but potential and 
ultrasoft loop momenta are still not separated. First, we note 
that quark lines can never be ultrasoft. Technically, this is 
so, because, since only $k_0$ can combine with a large $q_0$, 
spatial tadpoles always result, if one tries to make quark lines 
ultrasoft. As a consequence, one should consider all contributions 
in which all quark lines are potential and a (possibly empty) 
collection of (possibly disconnected) gluon lines is ultrasoft. 
When an ultrasoft gluon line with momentum $l$ connects to a 
quark line with loop momentum $k-l/2$ for the incoming and $k+l/2$
for the outgoing quark line, we can expand the quark-gluon vertex 
and quark propagator in $\vec{l}/{\vec{k}}\sim \sqrt{\hat{y}}$. 
This corresponds to a multipole expansion, as the wavelength of ultrasoft 
gluons is large as compared to separation of the quark-antiquark 
system in a coordinate space picture. After multipole expansion, 
all scales are separated and we arrive at a fully homogeneous 
threshold expansion. 

We have arrived at this prescription in a heuristic and descriptive 
way. We are presently not able to give a mathematical proof 
of our prescriptions similar to
what is available for off-shell limits \cite{Sm1} and leave 
a mathematical treatment to future publications. 
To complete such a proof for the threshold
expansion one needs to use the $\alpha$-representation and a subsequent
decomposition of the integration domain into appropriate sectors
which should be introduced, in the language of the $\alpha$-parameters, 
in a way analogous to our decomposition
into regions of large, potential, soft and ultrasoft momenta.
In fact, a proof of the asymptotic expansions for off-shell limits
\cite{Sm1} reduces to the analysis of the asymptotic behaviour of the
remainder, which has the structure of the $R$-operation, i.e. 
renormalization at the diagrammatic level. Therefore this proof has much in
common with standard proofs of the finiteness of the
renormalized Feynman integrals.

According to the definition of the asymptotic expansion,
its remainder possesses a given asymptotic behaviour
(a sufficiently fast decrease) once one keeps a sufficiently
large number of the terms of the expansion. As in the case of  
the limits for which explicit general formulae were known up to now
\cite{Go,Ch,Sm1,acvs},
it is possible to characterize the remainder of the threshold
expansion in terms of
the $R$-operation.
Since the product of any two (or more) subtraction operators
that are present in the prescriptions for the asymptotic expansions
is zero, we can rewrite the sum that in the right-hand side
of the expansion (\ref{AE}) as
\be
\sum_{\gm} {\cal M}_{\gm} = 1 - {\cal R} \,
\ee
where the operation ${\cal R}$ has the structure of the usual
(ultraviolet) $R$-operation and is given by the forest formula
\be
{\cal R} = \sum_{\cal F} \prod_{\gamma\in {\cal F}}  
\left( -{\cal M}_{\gm} \right) \, . 
\label{FF}
\ee
For off-shell limits, the operator ${\cal M}_{\gm}$ performs
Taylor expansion of the Feynman integral  $F_{\gm}$ in its small
masses and external momenta and inserts the resulting polynomial
in the reduced diagram $F_{\Gm/\gm}$. In our case, the Taylor 
expansions appropriate to a given region should be implied.  
In (\ref{FF}) the sum is  in forests (i.e. subsets of non-overlapping
subgraphs) composed of subgraphs that are involved in the prescriptions
for the asymptotic expansion in the given limit.
Therefore, we can represent an initial Feynman integral as
\be
F_{\Gm}= \sum_{\gm} {\cal M}_{\gm} F_{\Gm}  + {\cal R}F_{\Gm} \, .
\ee

The remainder is constructed in such
a way that it is finite (if the original diagram
is finite) and possesses a desired estimate in the given limit.
The order of the expansion is implicitly hidden in
degrees of the subtractions operators involved. 
From the mathematical point of view,
the subtractions performed by the operators ${\cal M}_{\gm}$
corresponding to subgraphs $\gm \neq \Gm$
remove divergences generated by the naive expansion (i.e. for
$\gm =\Gm$) of the
integrand in the expansion parameter. This naive expansion in
turn removes divergences that could appear due to the subtractions
in the subgraphs.

Let us illustrate the above points through our examples.
In example~1, we have the following expression for the remainder:
\be
{\cal R}I_1\equiv \int [dk]
\left(1- {\cal T}_y^N \right) \left( 1- {\cal T}_{k_0^2}^N \right)
\frac{1}{(k^2+q\cdot k-y) (k^2-q\cdot k-y) 
(k-p)^2} \, .
\label{ex1_rem}
\ee
Here the superscript $N$ denotes the order of the Taylor expansion. 
It turns out that the remainder possesses a necessary asymptotic behaviour,
provided the degree of the subtraction operators is chosen sufficiently
large, and does not have divergences that were not present from the
very beginning. Consider first the region of small $k$. Then the operator
${\cal T}_y^N$ is dangerous, in the sense that it generates an infrared 
divergence
that appears when $y \to 0$. However, the second factor
$(1- {\cal T}_{k_0^2}^N)$ kills this divergence because
it either provides additional factors of $k$, or explicit powers
of $y$ (which effectively reduce the order of expansion in $y$
in the first factor). 
Similarly, in the region of large $k$, the operator
${\cal T}_{k_0^2}^N$ is dangerous because it generates positive powers
of $k$. This time, the first factor $(1- {\cal T}_y^N)$
improves the convergence of the whole integral.

Consider example~2. 
As a remainder we now have ${\cal R}F_{\Gm}$ where
\be
{\cal R} = (1-{\cal M}_{\Gm}) (1-{\cal M}_{h-p}-{\cal M}_{p-h})
(1-{\cal M}_{p-p}-{\cal M}_{p-us}) \, .
\label{rem4}
\ee
The operators ${\cal M}_{h-p}$, ${\cal M}_{p-h}$, ${\cal M}_{p-p}$ and 
${\cal M}_{p-us}$ have to remove infrared divergences
generated by the action of the operator 
${\cal M}_{\Gm}={\cal M}_{h-h}={\cal T}_y$. 
Consider, for example, the region where $k$ is small while $l$ is
non-zero. Then one can pick up the bracket $(1-{\cal M}_{h-p})$ which makes
the term with  ${\cal M}_{\Gm}F_{\Gm}$ convergent.
In the region where both $k$ and $l$ are small, this is the operator
${\cal M}_{p-p}$  that provides convergence that could be spoiled
by the operator ${\cal M}_{\Gm}$,
and the operator ${\cal M}_{p-us}$ that kills the divergence at $k,l \to 0$
caused by ${\cal M}_{h-p}$ and ${\cal M}_{p-h}$. 

As in the case of known limits of momenta
and masses, we have, in the right hand-side of (\ref{ex1_rem}) and 
(\ref{rem4}), and (\ref{exp_sm_la}) and (\ref{exp_s_pus}),  
an interplay between ultraviolet and infrared 
divergences which manifest themselves as poles in $\eps$. However,
the naive expansion (in $y$) does not always generate  
additional poles --- see, e.g., example~1. Still the naive Taylor 
expansion part taken alone does not provide a correct 
answer to the problem and should be accompanied by 
extra contributions.\\

{\large\bf \noindent 4. Two-loop vertex integrals}\\ 

\begin{figure}[t]
   \vspace{-4.7cm}
   \epsfysize=30cm
   \epsfxsize=20cm
   \centerline{\epsffile{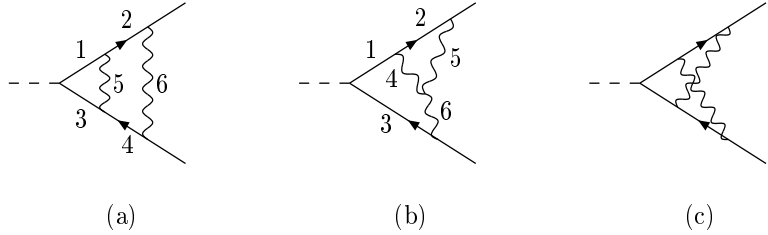}}
   \vspace*{-20.3cm}
\caption[dummy]{\small Examples of two-loop three-point graphs. 
Solid (wavy) lines denote massive (massless) lines. \label{fig4}}
\end{figure}

\noindent In this section we discuss and present 
results on the threshold expansion of two-loop 3-point functions, which, 
to our knowledge can not be obtained from expansion of an exact result. 
The three-point functions are also the non-trivial building blocks for 
many physical applications. Here we consider the abelian planar and 
non-abelian graphs in Fig.~\ref{fig4}a and b. The non-planar graphs 
(Fig.~\ref{fig4}c) will be discussed together with first applications 
in \cite{BSS}.\\ 

{\em The abelian planar graph.} Define the one-loop subgraphs 
$\gamma_1=\{1,3,5\}$, $\gamma_2=\{2,4,5,6\}$ and $\gamma_3=\{1,2,3,4,6\}$. 
(The numbering of lines refers to Fig.~\ref{fig4}a.) The unexpanded 
integral is given by
\be
I_{\rm PL} 
\equiv\int \!\frac{[dk] [dl]}{(k^2+q\cdot k-y) (k^2-q\cdot k-y)
(l^2+q\cdot l-y) (l^2-q\cdot l-y) (k-l)^2 (l-p)^2}, 
\label{planar}
\ee
where the kinematic variables are defined as in example 1 above. 
Inspecting all regions that contribute to the threshold expansion, 
we find only six of them non-vanishing. Together with their 
leading contribution (up to logarithms of $\hat{y}$) to $q^4 I_{\rm PL}$, 
they are:
\bea
\mbox{h-h:} && 1 \nonumber\\
\mbox{h-p, with $\gamma_1$ hard:} && 1/\sqrt{\hat{y}} \nonumber\\
\mbox{h-p, with $\gamma_2$ hard:} && \sqrt{\hat{y}} \\
\mbox{p-p:} && 1/\hat{y} \nonumber\\
\mbox{p-s, with $\gamma_2$ soft:} && 1/\sqrt{\hat{y}} \nonumber\\
\mbox{p-us, with $\gamma_3$ potential:} && 1/\sqrt{\hat{y}}. \nonumber
\eea
Let us emphasize that the scalings given here follow from power 
counting alone, because in each region the integrand is homogeneous 
in the expansion parameter.

The calculation of the various contributions is straightforward in the 
sense that all non-trivial aspects of the calculation have been met 
in the examples discussed in detail in Sect.~2. For instance,  
after partial fractioning, the integrals needed to 
evaluate the h-h region reduce to $J_\pm$ of (\ref{Jints}) (with 
additional scalar products in the numerator) and 
the complexity of integrals is essentially the same as for the 2-point 
function of example 2. The leading contribution close to threshold 
comes from the p-p region and corresponds to the double Coulomb 
exchange. Let us consider in more detail the fifth contribution, when 
the box subgraph is soft and loop momentum $k$ that flows through 
lines 1, 3 and 5 is potential. First, when we take the integral over the 
other loop momentum $l$ first, the result is identical to the 
soft contribution to the box subgraph in example 3 of Sect.~2, except 
that the difference of external relative momenta $p-p'$ is 
replaced by the difference of the external momenta of the box subgraph 
$p+k-p=k$. The momentum dependence of the box subgraph is 
$1/(\vec{k}^2)^{1+\epsilon}$, which can be interpreted as a 
non-local vertex. Second, using $p\cdot q=0$, it is easy to see 
that in every region the expansion parameter is $\hat{y}$, except 
when one momentum is soft, in which case the expansion parameter 
is $\sqrt{\hat{y}}$. However, for the box subgraph the symmetry between 
the upper and lower quark line eliminates all odd terms in 
$\sqrt{\hat{y}}$ and subleading contributions are suppressed by 
integer powers of $\hat{y}$. 

The integral $I_{\rm PL}$ is infrared divergent. Collecting all terms 
up to order $\hat{y}^0$ close to threshold, we obtain
\bea
\label{planar3pt}
\left(\frac{q^2}{4}\right)^{\!2+2\epsilon} \!\!I_{\rm PL} &=& 
\frac{1}{\epsilon^2} \left[\frac{\pi^2}{128\hat{y}}
-\frac{\pi}{16\sqrt{\hat{y}}}+\frac{1}{8}\right] 
+\frac{1}{\epsilon} \left[-\frac{\pi^2}{64\hat{y}}\,\ln(16\hat{y})+
\frac{\pi}{8\sqrt{\hat{y}}} \left(2-\ln 2\right) - 
\frac{1}{4}\right] \nonumber \\
&&\hspace*{-2.5cm}
+\,\frac{\pi^2}{64\hat{y}} \left(\ln^2(16\hat{y})+\frac{7\pi^2}{12}
\right) + \frac{\pi}{8\sqrt{\hat{y}}} \left(\ln^2(16\hat{y})+(3\ln2-4)\,
\ln(16\hat{y}) + \ln^2 2-2\ln 2+\frac{\pi^2}{4}\right) \nonumber\\
&&\hspace*{-2.5cm}
-\,\frac{29\pi^2}{48}-\frac{3}{2}+ 
{\cal O}(\hat{y}^{1/2}).
\eea
The calculation of higher-order terms can be automated on a computer. 
As a check of this result, we have calculated the exact double pole 
part of the integral,
\be
\left(\frac{q^2}{4}\right)^{\!2+2\epsilon} \!\!I_{\rm PL\,|\,
\frac{1}{\epsilon^2}} = \frac{1}{32\epsilon^2 \hat{y}}
\left(\frac{\pi}{2}-\arctan\sqrt{4\hat{y}}\,\right)^2,
\ee 
and the first three terms in the expansion agree with the double pole part 
of (\ref{planar3pt}).\\

{\em The non-abelian graph.}  For the diagram of Fig.~\ref{fig4}b, 
we define 
the one-loop subgraphs $\gamma_1=\{2,4,5\}$, $\gamma_2=\{1,3,4,6\}$ 
and $\gamma_3=\{1,2,3,5,6\}$. The unexpanded 
integral is given by
\be
I_{\rm NA} 
\equiv\int \!\frac{[dk] [dl]}{(k^2+q\cdot k-y) (k^2-q\cdot k-y)
(l^2+q\cdot l-y) (k-l)^2 (k-p)^2 (l-p)^2}. 
\label{non-abelian}
\ee
Inspecting all regions that contribute 
to the threshold expansion, 
we find only four of them non-vanishing. Together with their 
leading contribution (up to logarithms of $\hat{y}$) to $q^4 I_{\rm NA}$, 
they are:
\bea
\mbox{h-h:} && 1 \nonumber\\
\mbox{h-p, with $\gamma_1$ hard:} && 1/\sqrt{\hat{y}} \\
\mbox{p-s, with $\gamma_1$ soft:} && 1/\hat{y} \nonumber \\
\mbox{p-us, with $\gamma_3$ potential:} && 1/\sqrt{\hat{y}}. \nonumber
\eea
The p-p region that scales as $1/\hat{y}$ does not contribute here, 
because the poles in $l_0$ lie on one side of the real axis only. 
Hence, the $l_0$-integral vanishes. The potential-soft region 
is algebraically most complicated. Since the soft vertex subgraph 
$\gamma_1$ is less symmetric than the box subgraph of the abelian 
planar diagram, the expansion in this region runs in $\sqrt{\hat{y}}$ 
and it is necessary to compute the first two subleading terms in the 
expansion in this region to obtain an accuracy ${\cal O}(\hat{y}^0)$ 
at threshold. The integral $I_{\rm NA}$ is infrared 
divergent. Collecting all terms 
up to order $\hat{y}^0$ close to threshold, we obtain
\bea
\label{nonabelian3pt}
\left(\frac{q^2}{4}\right)^{\!2+2\epsilon} \!\!I_{\rm NA} &=& 
\frac{1}{\epsilon^2} \left[-\frac{\pi}{32\sqrt{\hat{y}}}+\frac{1}{8}\right]  
+\frac{1}{\epsilon} \left[\frac{\pi^2}{64\hat{y}}-\frac{\pi^2}{16}
-\frac{1}{4}\right] -\,\frac{\pi^2}{32\hat{y}}\left(\ln(16\hat{y})+2\right)
\\
&&\hspace*{-2cm} 
\,+\frac{\pi}{32\sqrt{\hat{y}}} \left(\ln^2(16\hat{y})+8\ln(32\hat{y})
+\frac{5\pi^2}{6}\right)+\frac{\pi^2}{8}\ln(16\hat{y})
-\frac{23\pi^2}{48}-\frac{3}{2} 
+{\cal O}(\hat{y}^{1/2}).\nonumber
\eea
As a check of this result, we have calculated the exact double pole 
part of the integral, 
\be
\left(\frac{q^2}{4}\right)^{\!2+2\epsilon} \!\!I_{\rm NA\,|\,
\frac{1}{\epsilon^2}} = 
-\frac{1}{16\epsilon^2 \sqrt{\hat{y}}}
\frac{1}{1+4\hat{y}}
\left(\frac{\pi}{2}-\arctan\sqrt{4\hat{y}}\,\right),
\ee 
and the first two terms in the expansion agree with the double pole part 
of (\ref{nonabelian3pt}).\\

We close this section with some observations on the non-planar 
diagram, Fig.~\ref{fig4}c, postponing a detailed discussion to 
\cite{BSS}. The non-vanishing regions are the same 
as for the non-abelian diagram, with the difference that the hard-potential  
contribution scales as $\sqrt{\hat{y}}$ in leading order and the 
potential-soft region as $1/\sqrt{\hat{y}}$ as for the planar abelian 
diagram. (The p-p 
region vanishes again, because all poles in one of the zero-component 
integrations lie on one side of the real axis.) Any integral that 
contributes to the h-h region can be reduced to either $J_\pm$ or 
\bea
\label{Lints}
L_\pm(a_1,\ldots,a_5)\equiv && \nonumber\\
&&\hspace*{-2.8cm}
\int\!\frac{[dk] [dl]}{[-k^2]^{a_1} 
[-l^2]^{a_2} [-((k+l)^2+q\cdot (k+l))]^{a_3} [-(k^2+q\cdot k)]^{a_4} 
[-(l^2\pm q\cdot l)]^{a_5}}. 
\eea
With integration by parts we can reduce these integrals first to 
$L_+(0,0,a_3,a_4,a_5)=L_-(0,0,a_3,a_4,a_5)$ and then to a single 
integral such as $L_+(0,0,2,2,1)$. (The algorithm for $L_+$ has been used 
before in \cite{GRA,FT}, although no details have been given there.) 
Furthermore, the hard parts of all possible 2-loop 3-point 
graphs relevant to QCD can be reduced to $J_\pm$ or $L_\pm$ and, 
consequently, can be solved in the threshold expansion.\\ 

{\large\bf \noindent 5. Discussion}\\

\noindent The threshold expansion, which we 
proposed and illustrated in this paper, 
could be applied to two sets of problems. The first set consists of 
problems where it is desirable to have as many terms in the expansion 
as can possibly be calculated. The second consists of problems where 
the nature of the problem specifies that only a limited number of 
terms are needed.

The cross section for the production of heavy quarks in $e^+ e^-$ 
annihilation $\sigma_{e^+e^-}(q^2,m)$ at intermediate energies 
$\sqrt{q^2}$ is an important example of the first kind. Presently 
available techniques do not permit an analytic calculation of  
$\sigma_{e^+e^-}(q^2,m)$ at order $\alpha_s^2$. In \cite{CKS} the 
large-momentum and large-mass expansion together with convergence 
accelerating methods have been used to approximate 
$\sigma_{e^+e^-}(q^2,m)$ to high accuracy from expansions around 
$q^2=0$ and $q^2=\infty$. It is exactly the threshold point 
$q^2=4 m^2$ that is most critical in this approach. The threshold 
expansion, when computed to high order, provides us with 
an intermediate point, the expansion around which can probably be 
smoothly joined with the large-momentum and large-mass expansion 
in order to obtain an accurate result for $\sigma_{e^+e^-}(q^2,m)$ 
for any $q^2$ except very close to threshold, where a resummation 
of all corrections of form $\alpha_s^{0,1,2} (\alpha_s/\sqrt{\hat{y}})^n$ 
is necessary. (This resummation belongs to the second set of problems.)  
The analysis of the 2-loop 2-point function in Sect.~2 suggests that the 
threshold expansion for $\sigma_{e^+e^-}(q^2,m)$ 
converges when $q^2\in [0,8 m^2]$. Therefore 
we expect that a reasonably accurate approximation to 
$\sigma_{e^+e^-}(q^2,m)$ could be obtained for $q^2<6 m^2$ starting 
from the threshold expansion.

In this paper we have treated only the threshold expansion of loop 
integrals, but not of phase space integrals for real radiation. 
However, as in \cite{CKS} it seems advantageous to consider the 
three-loop vacuum polarisation of heavy quarks in the threshold 
expansion in order to avoid the separate calculation of virtual 
corrections and real radiation. Since $\sigma_{e^+e^-}(q^2,m)$ 
requires only the imaginary part of the two-point function, one can 
take advantage of this fact from the beginning. For example, the 
region where all three loop momenta are hard, which would be 
very difficult to calculate, is analytic in 
$\hat{y}$ and therefore does not contribute to the imaginary part. 
Other contributions reduce to two-loop diagrams (such as those of 
Sect.~4) or simple three-loop diagrams. 

The second set of problems involves `matching calculations'. In 
this case the quark mass $m$ is a large scale and the observable 
in question should be factorized into a short-distance contribution 
from the scale $m$ and a long-distance contribution which can be 
either perturbative or non-perturbative. Examples of this kind include 
bound state calculations in QED, quarkonium systems in QCD, but also 
the above-mentioned resummation of `Coulomb-enhanced' corrections 
in heavy quark production close to threshold. In order to obtain the 
matching coefficient for a given operator, a suitable (typically on-shell) 
Green function has to be expanded close to threshold to a certain order, 
which is determined by the structure and dimension of the operator.
As already mentioned in Sect.~3, the combinatorical structure of 
the separation of large (hard) and small momenta in the threshold 
expansion, which is manifest in (\ref{exp_sm_la}) for a given 
diagram $\Gamma$, translates into the statement of factorization 
\be 
G(m,p_i) = \sum_i C_i(m)\,G_{O_i}(p_i/m)
\ee
for the Green function $G$ to which $\Gamma$ contributes. Here $O_i$ is 
a local operator, inserted into $G$, and $C_i$ is its short-distance 
coefficient. This construction is completely equivalent to the 
construction of a non-relativistic effective theory \cite{CL} in 
dimensional regularization. In particular (\ref{exp_sm_la}) leads to 
the factorization prescription that for any diagram $\Gamma$ its 
contribution to the short-distance coefficient(s) (`hard part') is 
given by selecting only $\gamma=\Gamma$ in the sum over all 
subgraphs:  
\be
F_{\Gamma\,|\,\mbox{hard}} = {\cal T}_y F_{\Gamma}.
\ee
That is, the hard part is given by the `naive' Taylor expansion of 
$\Gamma$. Note that this is not identical to dropping all singular terms 
at threshold. For example, the abelian planar 3-point integral 
in Sect.~4 contains terms proportional to $\hat{y}^0$ both from 
the h-h and the p-p region. The contribution from the second region 
should not be considered as part of the short-distance coefficient. 
The non-relativistic approximation to the quark propagator in 
(\ref{softapprox}) does not coincide exactly with the standard 
convention. Let the gluon line in the triangle graph of Fig.~\ref{fig1} 
have momentum $k$ and the upper quark line have momentum $k+p_1$. 
In the standard convention one would introduce an external momentum 
$\tilde{p}=(\tilde{p}_0,\vec{p})$ with the non-relativistic 
on-shell condition 
$\tilde{p}_0=\vec{p}^{\,2}/(2 m)$. The non-relativistic propagator is 
then
\be
\frac{1}{2 m}\,\frac{1}{\tilde{p}_0+k_0-\frac{(\vec{p}+\vec{k})^2}{2 m}} 
= \frac{1}{-(\vec{k}+\vec{p})^2+q_0 k_0-y+\delta},
\ee
where the difference to the approximation used in this paper is 
$\delta=(q_0/2-m)\,k_0 = (\sqrt{m^2-y}-m)\,k_0$, which is smaller than all 
other terms in the denominator and can be expanded. Thus our 
small-momentum approximation is equivalent to a reparametrization of 
NRQED/NRQCD, where the large part of the massive 
line momentum is taken to be $q/2$ rather than $m$.

Once the hard momenta are factorized (`integrated out'), an interesting 
aspect of non-relativistic effective theory is that it does not 
lead to manifest power counting. This means that the matrix elements 
of operators are series in the small parameters, rather than being 
homogeneous, and their scaling can not be determined by counting a 
suitable dimensionful parameter. In the threshold expansion this is 
reflected by the fact that we had to divide the small momentum region 
further into potential, soft and ultrasoft (see (\ref{softterm})) in 
order to arrive at a homogeneous expansion with manifest power counting. 

Recently the possibility to factorize the small-momentum scales 
$m v$ and $m v^2$ ($\sqrt{y}$ and $y/q$ in our notation) has been 
investigated in a series of papers \cite{LAB,LM,GR,LS,PS}. Different 
rescalings were proposed in \cite{LM,GR} that succeed in making 
power counting manifest in one region, but fail in others. Thus 
while the rescaling of \cite{LM} is adequate for potential gluons 
($k_0\sim m v^2$, $\vec{k}\sim m v$), the rescaling of \cite{GR} is 
tuned for ultrasoft gluons ($k_0\sim \vec{k}\sim m v^2$) 
and the multipole expansion. Both 
rescalings are unified in the treatment of \cite{LS}, where 
different fields for (non-propagating, off-shell) potential gluons 
and (propagating) ultrasoft gluons are introduced. At least for diagrams, 
in which only one gluon (massless) line is present, this approach appears 
to be equivalent to ours, as can be seen by comparing the treatment 
of the small momentum contributions to the triangle graph in Sect.~2 
with the discussion in \cite{LS}. Furthermore, in the third line of 
Fig.~\ref{fig2} the first graph corresponds to the exchange of 
a potential gluon and the second to the exchange of a propagating 
gluon in the language of \cite{LS}. On the other hand, the existence  
of soft gluons, i.e. propagating gluons with energy and 
momentum of order $m v$, has not been considered in \cite{LM,GR,LS} 
and does not seem to fit into the rescaling schemes proposed there. 
Yet another approach has been taken in \cite{PS}, where a non-local 
effective Lagrangian is proposed that follows after integrating out soft 
gluons. (In practice, this has been done only for single gluon exchange, 
which gives rise to the standard Coulomb potential with relativistic 
corrections. The discussion of the soft box subgraph in Sects.~2 and 4 
suggests that the non-local Lagrangian can be extended to include 
the box graph contribution.) However, it has not yet been shown 
how to treat potential gluons in the approach of \cite{PS}. 

Labelle has first discussed the separation of soft/potential 
and ultrasoft photons and the multipole expansion in the context of
NRQED \cite{LAB}. It seems to us that Labelle's treatment of NRQED graphs 
is most closely related to our treatment of the small momentum regions 
in the threshold expansion. Technical differences arise in as much 
as he implies a cut-off regularization, while the simplicity of the 
threshold expansion discussed here is specific to dimensional regularization. 
In addition, \cite{LAB} starts with time-ordered perturbation theory 
and Coulomb gauge, while we emphasize that the power counting 
and approximations in a given region can be carried out before 
performing the integration over the zero-components of the loop momenta 
and the method works in any gauge. (Since we have considered scalar 
propagators only, the question of gauge does not arise. In a gauge 
theory Coulomb gauge is very convenient for the threshold expansion, 
because propagating gluons which cause most complications couple to 
quarks through vertices that are suppressed by at least a factor of 
$\sqrt{y}$.) In this paper, as in \cite{LAB} 
and in contrast to \cite{LM,GR,LS,PS},  
no attempt is made to write down an effective Lagrangian after 
integrating out soft or soft together with potential gluons. 
This may be considered 
as a drawback. On the other hand, as this Lagrangian would  
inevitably be non-local, its utility is not obvious to us, once manifest 
power counting can be achieved otherwise.

After this conceptual postludium we should emphasize that the main focus 
of this paper is on the ability to perform calculations. The threshold 
expansion method described here allows us to treat 2-loop 3-point 
functions and opens the door to threshold problems at 
next-to-next-to-leading order. A first realistic application will 
be treated in a companion paper \cite{BSS}.\\

\noindent {\bf Acknowledgements.} We thank A. Czarnecki, A. Davydychev 
and K. Melnikov for helpful discussions and providing 
checks on some of the integrals 
required for the hard-hard region. We are also grateful to 
A.V.~Kotikov and I.Z.~Rothstein for useful discussions and to 
W.~Beenakker and B.~Tausk for correspondence regarding the 
exact box graph. 
V.S. thanks the CERN Theory Group and the Universities of  
Leiden and Mainz for hospitality during the course of this work. 
Both, V.S. and M.B. thank the Theory Group at the 
University of Karlsruhe for hospitality while this work was completed. 
Thanks to G. Buchalla and K.~Chetyrkin for reading the manuscript.
The work of V.S. has been supported by the Russian Foundation for 
Basic Research, project 96--01--00726, and by INTAS, grant 93--0744.


\end{document}